\documentstyle[prl,aps,amssymb,multicol,epsfig]{revtex}

\input epsf

\newcommand{\Wmul}[1]{\begin{multicols}{#1}}
\newcommand{\Wmuldue}{\begin{multicols}{2}}
\newcommand{\Wemul}  {\end{multicols}}

\begin{document}

\preprint{}

\title{ An anomalous alloy: $Y_xSi_{1-x}$}                        

\author{V. ~Meregalli, M. ~Parrinello}                   

\address{Max-Planck-Institut f\"ur Festk\"orperforschung\\ 
  Heisenbergstrasse.~1, D-70569 Stuttgart, Germany} 

\date{\today}

\maketitle

\begin{abstract}
We study via density functional-based molecular dynamics the structural 
and dynamical properties of the rare earth silicon amorphous alloy 
$Y_xSi_{1-x}$ for $x=0.093$ and $x=0.156$ . The $Si$ network forms cavities 
in which a $Y^{3+}$ cation is entrapped. 
Its electrons are transferred to the $Si$ network 
and are located in the dangling bonds of the $Si$ atoms that line the $Y$ 
cavities. This leads to the presence of low coordinated $Si$ atoms that 
can be described as monovalent or divalent anions. 
For $x=0.156$, the cavities touch each other and share $Si$ atoms that 
have two dangling bonds. 
The vibrational spectrum is similar to that of 
amorphous $Si$. However, doping induces a shoulder at 70 cm$^{-1}$
and a pronounced peak at 180 cm$^{-1}$
due to low coordinated $Si$.
\end{abstract}

 \draft

\Wmuldue{}

 Rare earth doped silicon alloys have received considerable attention due
 to their intriguing properties and for their potential in the applied 
domain. They show an interesting metal-insulator transition
and their optical
conductivity exhibits apparent sum rule violations \cite{fh},
as in manganates and high $T_c$ compounds. 
In addition $Gd_xSi_{1-x}$ alloys have an enormous negative magneto resistance
\cite{fh2} 
and spin-glass behaviour has been reported \cite{fh3}. 
In spite of a large number of 
experimental investigations the structure of this and related compounds 
is still a matter of guesswork. For this reason we have undertaken 
here an {\it ab-initio} molecular dynamics investigation of 
the $Y_xSi_{1-x}$ alloy.
This is non-magnetic but has electronic structure properties 
that set it apart from all the other amorphous 
$Si$ alloys so far studied. 

In our simulation we followed the method of Car and Parrinello \cite{cpmd1} 
as implemented in the CPMD code \cite{cpmd2}. 
Local density approximation and norm-conserving Goedecker-type 
pseudopotentials have been used \cite{pp}, \cite{pp2} (for $Y$ the semi-core
4d states have been included in the valence). 
The simulation cell is a cube of length 11.216\ \AA{} containing a total of
64 atoms.
Periodic boundary conditions are applied.
The concentrations $x=0.093$ ($58Si+6Y$) and $x=0.156$
($54Si+10Y$) have been studied. The Car-Parrinello equations of motion 
with a fictitious electronic mass of 400 a.u. were integrated using a 
time step of 3 a.u. (0.072 fs). In order to generate the amorphous 
structure we used the following procedure. We start from a cubic lattice 
with the $Y$ atoms placed randomly, but as far as possible from one another. 
Such an unlikely configuration is rapidly destroyed and leads 
to fast melting. The system is kept in this liquid state at the temperature 
of 1650 K for about 0.35 ps, in order to lose memory of the initial
 condition. Then the temperature is brought down to 300 K in a time span
of 0.3 ps. 

\begin{figure}[tbp]
  \epsfxsize=7.0cm \epsfbox{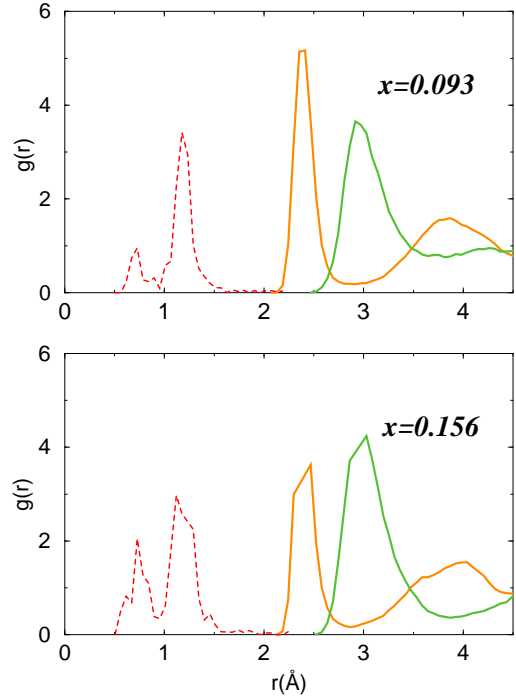}
  \caption{
    $Si-Si$ (orange line),  $Si-Y$ (green line) and $Si-WFCs$ (red line)
    pair correlation functions for both the concentrations.
    The data have been obtained by averaging over all the configurations
     of the  molecular dynamics simulations. The $Si-WFCs$ correlation
     function has been scaled by a factor of 3.
    }
  \label{gr}
\end{figure}

Although the annealing rate is rather fast, past experience in 
semiconducting alloys has shown that the main features of the amorphous 
structure are well reproduced \cite{am} . 
The system was finally allowed to evolve at 300 K for 10 ps
without ionic temperature control. The system is a metal in the
 liquid state and a small gap semiconductor when amorphous at room temperature. 
In the
 annealing part of the run we controlled the fictitious 
electronic kinetic energy 
drift by periodic quenches to the Born Oppenheimer surfaces. Instead a 
Nose' thermostat was used for the electrons in the last 10 ps. The thermostat
 characteristic frequency was 1500 cm$^{-1}$ and the target electronic 
fictitious kinetic energy was set according to the prescription of reference \cite{ke}. 
The vibrational spectrum and other statistical averages were 
evaluated during the last 10 ps. 

In Fig.\ \ref{gr}  we report the measured correlation functions $g_{SiSi}$ and 
$g_{YSi}$. Due to the small number of atoms the $g_{YY}$'s 
are very noisy and are not reported in order not to clutter the figure. 
However they show a strong tendency for the $Y$ to repel each other. 
From the correlation functions as well as a visual inspection of the 
structure we could infer 
that $Si$ atoms tend to form 
a tetrahedrally bonded amorphous network as in pure $a-Si$. 
However $Y$ atoms are not part of the $Si$ amorphous network but sit 
at the centre of cavities. In order to accommodate these cavities the $Si$ 
network has to generate a number of under-coordinated $Si$ atoms in the 
proximity of the $Y$ impurities. 

A very valuable help in understanding the electronic properties of 
amorphous structures 
is provided by the maximally localised Wannier functions \cite{wc1} 
and their centres, as discussed in ref.\ \cite{wc} for the case of pure $a-Si$. 
Wannier functions are obtained from the Kohn and Sham orbitals by 
performing a unitary transformation that minimises the 
average orbital spread. 
The resulting orbitals capture the chemical nature of the bond. 
An even simpler and more vivid picture is obtained by considering only 
their centroids.

This is demonstrated in Fig.\ \ref{gr} where we plot the $Si$ Wannier centre correlation 
functions. The fist peak is split into two. The higher peak is centred 
at half of the $Si-Si$  bond distance and clearly reflects the formation 
of the covalent bond. There is, however, a very sizable peak at shorter 
distances which is due to the presence of dangling bonds, whose height 
increases with $Y$ concentration. Since these peaks are well separated we
 can classify the Wannier centre and the corresponding Wannier functions 
into ``covalent'' and ``dangling''. 
The different nature of these classes 
of functions is reflected also in their spread distribution, which is 
bimodal with peaks centred at $\sim$ 1.5\ \AA{} and $\sim$ 1.85\ \AA{} 
for ``covalent'' and ``dangling'' bonds respectively. 

As in ref.\ \cite{wc} 
we can identify the covalently bonded $Si-Si$ atoms as those pairs of 
atoms which share  ``covalent'' Wannier functions. 
Using this criterion, which is based on the chemistry of the system and not 
simply on a geometric distance consideration, we find that in the 
$Si$ network there are twofold, threefold  and fourfold-coordinated atoms. 
As a function of the concentration, due to the need 
to accommodate more $Y$ impurities,
the number of threefold-coordinated silicons increases noticeably 
(from $20 \% $to $40\%$) while that of 
fourfold-coordinated atoms decreases 
(from $70\%$ to $50\%$).
We also noticed an increase of the twofold-coordinated
silicons; however, due to the limited statistics we do not attempt to 
define here the probability of their occurrence.

\begin{figure}[tbp]
   \epsfxsize=9.0cm \epsfbox{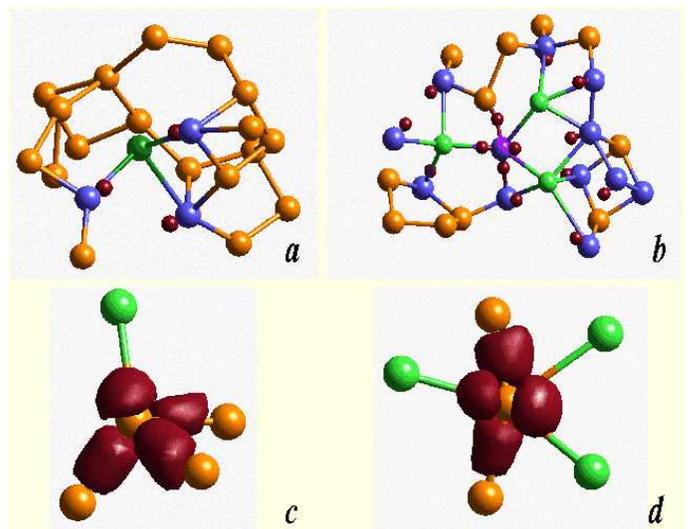}
  \caption{
     Snapshots of two configurations of the molecular dynamics simulation
     for both concentrations
     $x=0.093$ (a) and $x=0.156$ (b).
     Fourfold-coordinated silicon atoms are represented in orange,
     threefold-coordinated in blue, twofold-coordinated in purple,
     yttrium atoms in green and lone-pair Wannier centres in red.
     For clarity only the lone-pair Wannier centres have been plotted.
     Below each configuration we have plotted the Wannier function density
     $\rho_n({\bf r})=\left |w_n({\bf r})\right|^2$,
     for one of the threefold-coordinated atoms above (c)
     and for the twofold-coordinated atom above (d).
    }
  \label{wfc}
\end{figure}

In Fig.\ \ref{wfc}  we show two typical environments for the $Y$ atoms. 
At the higher concentration the $Y$ atoms have to share some of their 
first neighbours. These shared $Si$ atoms tend to be twofold coordinated 
as shown in the picture. 
Finally the Wannier analysis shows clearly that $Y$ has lost all 
its valence electrons and is in the oxidation state  3+ . 
Thus the picture that emerges from our simulation is one
in which the $Y$ atoms donate their valence electrons to the $Si$ sub-lattice.
These electrons are then stored in the $Si$ network and localised in the 
doubly occupied dangling bonds of the threefold and twofold coordinated $Si$ 
atoms. These atoms are therefore negatively charged and can be thought of as 
being a realization of the species $Si^-$ and $Si^{--}$ .
These species tend to cluster around the $Y$ cation, not only to form the 
cavity in which the impurity sits but also to ensure local charge neutrality. 
As such they are an integral part of the structure. The silicon dangling bonds 
provide the ligand field to the metal ion. 

\begin{figure}[tbp]
   \epsfxsize=7.0cm \epsfbox{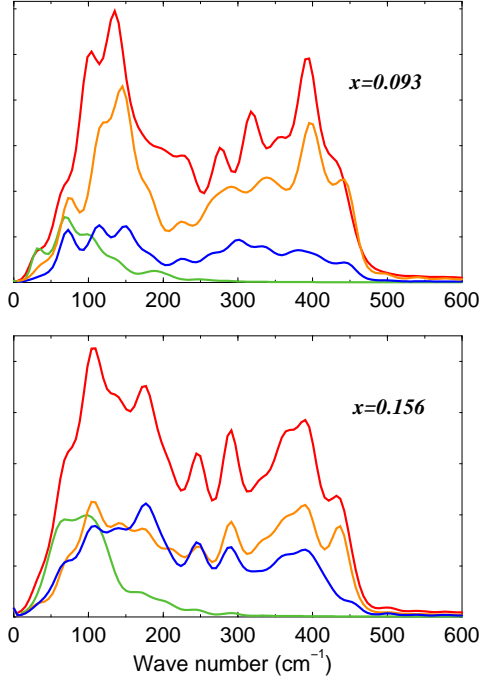}
  \caption{
    Total vibrational spectrum (red line) together with its contributions
     from $Y$ (green line), fourfold-coordinated $Si$ (orange line) and
    threefold-coordinated $Si$ (blue line), for $x=0.093$ (upper panel) and
    $x=0.156$ (lower panel).
    }
  \label{vs}
\end{figure}

In Fig.\ \ref{vs} we show the vibrational density of states analysed in the 
contributions from the species $Y$, $Si$ and $Si^-$.
The general appearance of the spectrum is very similar to that of pure $a-Si$.
However, we note a shoulder $\sim$ 70 cm$^{-1}$ which is due to the
$Y$ atoms; we also note that, especially for $x=0.156$, 
the $Si^-$ make an important
contribution and together with the $Y$ alter the relative intensities
of the peaks. Since $Si^-$ is expected to have a higher infra-red activity
than $Si$ we can predict that the infra-red spectrum will show an enhancement of
the 180 cm$^{-1}$ peak, and a $Y$ related structure at
70 cm$^{-1}$. The intensities of either peak should increase as a 
function of $x$.

For a selected small number of configurations we evaluated the HOMO-LUMO gap.
This shows large oscillations in the range 0.2 to 0.45 eV.

\begin{figure}[tbp]
  \epsfxsize=7.0cm \epsfbox{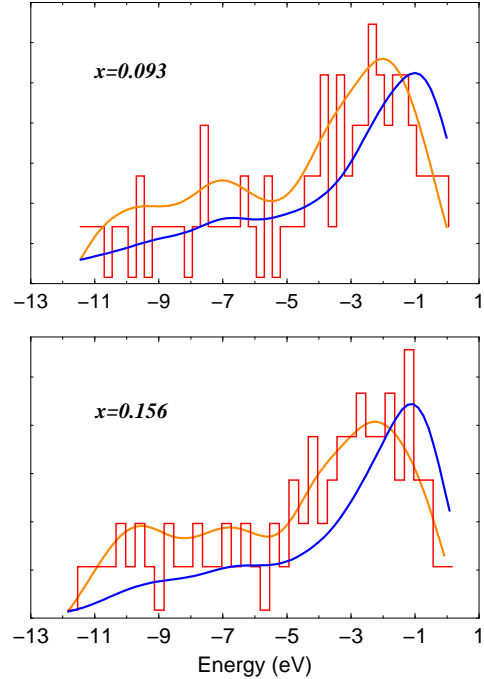}
  \caption{
    Electronic density of states (red line)
    and its projection on the covalent (orange line)
    and dangling (blue line) Wannier functions.
    Note that the projection are not normalised and the Fermi
    level defines the zero of the energy.
    }
  \label{wf}
\end{figure}

In Fig.\ \ref{wf} we show the total electronic density of states (DOS) 
and its projection on the covalent and dangling Wannier functions. 
These show clearly that the dangling states dominate the electronic 
DOS close to the Fermi level.

These two facts are not in contradiction to a picture in which the
conductivity takes place via an hopping mechanism as suggested by 
the experiments.

In conclusion, we have shown that $Y_xSi_{1-x}$ alloys at low concentrations
have a very peculiar electronic structure. The $Y$ donates its valence 
electrons to the $Si$  network and is trapped in cavities. 
These electrons are trapped in doubly occupied $Si$ dangling bonds. 
These defect states can be observed via infra-red spectroscopy.  
Our study can be a stepping stone for the more complex case of the magnetic 
$4d$ alloys. In fact it is likely that many of the features found here will also be present in the case of $Gd$ and contribute to its unusual properties.
We could speculate 
that as the $Y$ concentration increases the $Y$ filled cavities interact 
more and more until the tetrahedrically bonded network of $Si$ is 
disrupted and local coordinations like those observed in the 
stoichiometric compound $YSi_2$ will dominate the local structure. 
We plan to address these issues in our forthcoming work.

We thank  M.~Bernasconi, P.~L.~Silvestrelli and
F.~Hellman for stimulating discussions.

\Wemul{}

\end{document}